\documentclass[aps,prl,twocolumn,showpacs,superscriptaddress]{revtex4-1}

\usepackage{graphicx}
\usepackage{amssymb}
\begin{document}


\title{Ferromagnetic Spin Fluctuation and Unconventional Superconductivity in Rb$_{2}$Cr$_{3}$As$_{3}$ \\revealed by $^{75}$As NMR and NQR}


\author{J. Yang}
\affiliation
{Beijing National Laboratory for Condensed Matter Physics and Institute of Physics,
\\Chinese Academy of Sciences, Beijing 100190, P. R. China}

\author{Z. T. Tang}
\affiliation
{Department of Physics and State Key Lab of Silicon Materials, Zhejiang University, Hangzhou 310027, China}


\author{G. H. Cao}
\affiliation
{Department of Physics and State Key Lab of Silicon Materials, Zhejiang University, Hangzhou 310027, China}
\affiliation
{Collaborative Innovation Centre of Advanced Microstructures,  Nanjing 210093, China}

\author{Guo-qing Zheng}
\email
{gqzheng123@gmail.com}
\affiliation
{Beijing National Laboratory for Condensed Matter Physics and Institute of Physics,
\\Chinese Academy of Sciences, Beijing 100190, P. R. China}
\affiliation
{Department of Physics, Okayama University, Okayama 700-8530, Japan}


\date{\today}

\begin{abstract}

We report $^{75}$As nuclear magnetic resonance (NMR) and nuclear quadrupole resonance (NQR) studies on the  superconductor Rb$_{2}$Cr$_{3}$As$_{3}$ with a quasi one-dimensional crystal structure.
Below $T\sim$ 100 K, the spin-lattice relaxation rate (1/$T_{1}$) divided by temperature, 1/$T_{1}T$, increases upon cooling down to $T_{\rm c}$ = 4.8 K, showing a Curie-Weiss-like temperature dependence. The Knight shift also increases with decreasing temperature. These results suggest   ferromagnetic spin fluctuation. In the superconducting state, 1/$T_{1}$ decreases rapidly below $T_{\text{c}}$ without a  Hebel-Slichter  peak, and follows a $T^5$ variation below $T\sim$ 3 K, which point to unconventional superconductivity with point nodes in the gap function.

\end{abstract}


\maketitle

Unconventional superconductivity has been found in copper oxides~\cite{cuprates}, hydro cobalt-oxides~\cite{Takada} 
and
iron pnictides~\cite{Hosono}. In these materials that contain transition-metal elements,  unpaired 3d electrons are strongly correlated, which is believed by many  to be responsible for the unconventional nature of superconductivity.
Therefore, searching for unconventional superconductivity in other 3d-electron   materials has become an important topic.
 Indeed, a new family of Cr-based superconductors
was recently found and has received extensive attention.
By applying external pressures, CrAs  exhibits superconductivity with transition temperature $T_{\text{c}}$ $\thicksim$ 2 K on the border of a helical antiferromagnetic order~\cite{CrAs_IOP,CrAs_Japan}, and  unconventional electron pairing was  proposed in connection with a magnetic quantum critical point (QCP)~\cite{CrAs_IOP}.

More recently, A$_{2}$Cr$_{3}$As$_{3}$ (A = K, Rb, Cs)
family was discovered to be suerconducting with $T_{\text{c}}$ up to 6.1 K at ambient pressure~\cite{K,Rb,Cs}.
A$_{2}$Cr$_{3}$As$_{3}$ has a quasi one-dimensional (1D) crystal structure, which consists of infinite (Cr$_{3}$As$_{3}$)$^{2-}$ chains and A$^{+}$ as intercalated ions. Band structure calculations indicate that the electronic states at the Fermi level are dominated by Cr orbitals ~\cite{cao_c}, and that the Fermi surface consists of a three-dimensional (3D) surface and two quasi-1D sheets~\cite{cao_c,Wu}.
The 3D Fermi surface has a predominant contribution to the density of states (DOS) at the Fermi level~\cite{cao_c,private}.
The upper critical field $H_{c2}$ just below $T_{c}$ for (K,Rb)$_{2}$Cr$_{3}$As$_{3}$  shows a small   anisotropy~\cite{anisotropy,Hc2,Canfield}.
Nuclear quadrupole resonance (NQR)  measurement in K$_{2}$Cr$_{3}$As$_{3}$ found no coherence (Hebel-Slichter) peak in the temperature ($T$) dependence of the spin lattice relaxation rate ($1/T_1$) just below $T_{c}$~\cite{Imai}.
These results suggest unconventional superconductivity, which was echoed by other subsequent measurements ~\cite{Yuan_HQ,usr}. 
Theoretically, ferromagnetic spin fluctuations due to Cr-Cr direct interaction were suggested in view of the calculated electronic band structure and the peculiar crystal structure~\cite{cao_c,Hu_jiangping}, and  unconventional pairing states have been proposed ~\cite{zhang_fuchun,Wu}.

In this Letter, we report the first data set of the Knight shift in Rb$_{2}$Cr$_{3}$As$_{3}$  ($T_{\rm c}$ = 4.8 K) by $^{75}$As nuclear magnetic resonance (NMR), as well as the spin lattice relaxation via NQR measurements. 
In the normal state, 1/$T_{1}T$ is a constant above $T$=200 K as expected for a conventional metal, but is increased strongly upon cooling below 100 K, which indicates that  spin fluctuations develop at low temperatures. Our result is different from the previous work for K$_{2}$Cr$_{3}$As$_{3}$ where 1/$T_{1}T$ was reported to obey a power-law relation of $T^{-0.25}$ up to room temperature  ~\cite{Imai}. 
We find that the increase of 1/$T_{1}T$  can be  fitted by a  Curie-Weiss function. 
The Knight shift, which measures the spin susceptibility at the momentum $\emph{\textbf{q}}$ = 0, also increases with decreasing $T$ below $T\sim$ 100 K. These results are consistent with 3D  ferromagnetic spin fluctuation~\cite{Moriya}.  
In the superconducting state, no Hebel-Slichter coherence peak was found, as in K$_{2}$Cr$_{3}$As$_{3}$ ~\cite{Imai}. Furthermore, we find that 1/$T_{1}$ follows a $T^5$ variation below $T\sim$ 3 K.
Our results suggest    unconventional  superconductivity with point nodes in the gap function.


The Rb$_{2}$Cr$_{3}$As$_{3}$ polycrystalline sample used in this study  was prepared by solid state reaction method, as reported in Ref.~\cite{Rb}. For NMR/NQR measurements, we crushed the samples into powders and sealed the specimen into an epoxy (stycast 1266) tube. All operations were performed in a Ar-filled glove box to protect the samples against air and water vapor.
The  $^{75}$As nucleus has a nuclear spin \textit{I} = 3/2 and the nuclear gyromagnetic ratio $\gamma$ = 7.2919 MHz/T. The $^{75}$As NMR and NQR measurements were carried out by using a phase coherent spectrometer. The NMR and NQR spectra were obtained by scanning the RF frequency and integrating the spin echo at a fixed magnetic field $H_{0}$ = 11.9979 T and 0 T, respectively. The   $1/T_{1}$ was measured by the saturation-recovery method, 
and  determined by fitting  the nuclear magnetization   to $1-M(t)/M(\infty) = exp(-3t/T_{1})$, where $M(t)$ is the nuclear magnetization at time $t$ after the single saturation pulse \cite{Supple}.


\begin{figure}[tbp]
\includegraphics[width=8cm]{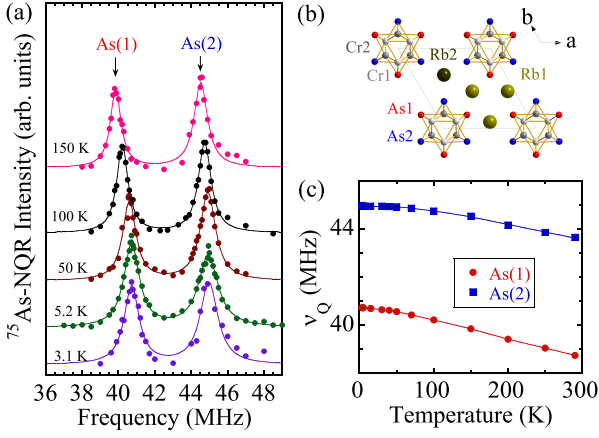}
\caption{(Color online). (a) $^{75}$As NQR spectra of Rb$_{2}$Cr$_{3}$As$_{3}$. The solid curves are Lorentzian function fittings with area ratio 1:1 of the two peaks. (b) The crystal structure  viewed from \emph{c}-axis direction. 
(c) The $T$-dependence of the $^{75}$As NQR frequency $\nu_{Q}$. The solid curves are guides for the eyes.
\label{NQR}}
\end{figure}

Figure.~\ref{NQR} shows the $^{75}$As NQR spectra which consist of two transition peaks. At all temperatures, each peak can be fitted by a Lorentz function, and the area ratio of the two peaks is 1 : 1.
  Theoretically, $^{75}$As NQR has only one transition ($m$ = $\pm$1/2 $\leftrightarrow$  $\pm$3/2) line. However, as shown in Fig.~\ref{NQR} (b), there are two different crystallographic As sites with atomic ratio As1 : As2 = 1:1.
Therefore, the observed two NQR peaks are from the two different As sites, but
 the correspondence between the two NQR transition lines and crystallographic sites is  unknown at the moment. 
In the following, we use As(1) to denote  the lower frequency
peak and As(2) the higher frequency peak for convenience.
Figure~\ref{NQR} (c) shows the obtained $T$-dependence of the $^{75}$As NQR frequency $\nu_{Q}$, which  increases gradually with decreasing $T$ but is almost  unchanged below $T$ = 50 K. 
 The well-resolved  NQR spectra  proved good quality of the sample. Note that in the previous report for K$_{2}$Cr$_{3}$As$_{3}$~\cite{Imai}, a third NQR peak was observed and ascribed to K-deficiency. In this sense, our spectra may indicate that  our sample is more stoichiometric.

\begin{figure}[tbp]
\includegraphics[width=8cm]{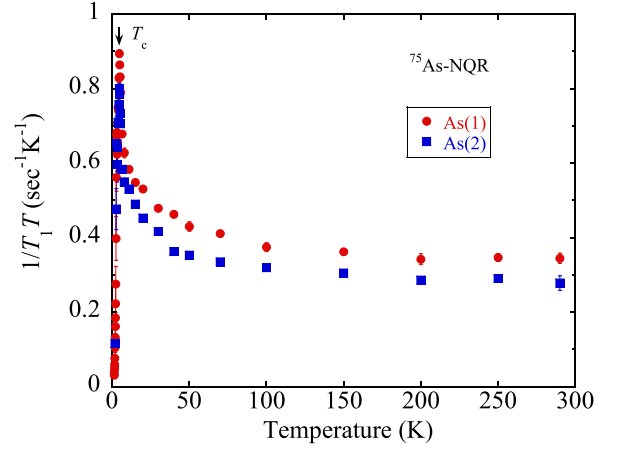}
\caption{(Color online). The temperature dependence of the $^{75}$As NQR spin-lattice relaxation rate divided by temperature (1/$T_{1}T$).
 \label{T1T}}
\end{figure}

Figure~\ref{T1T} shows the result of 
1/$T_{1}T$ measured via $^{75}$As NQR at zero magnetic field. The 1/$T_{1}T$ for the  both As sites shows a similar behavior. At $T$ $\geq$ 200 K,  1/$T_{1}T$ is a constant, as expected for a non-correlated conventional metal.
 However, strong enhancement of $1/T_{1}T$ upon cooling is found below 100 K and down to $T_{\text{c}}$ = 4.8 K. Below $T_{\text{c}}$, $1/T_{1}T$ drop rapidly due to the opening of a superconducting gap.  The $1/T_{1}T$ probes the imaginary part of the transverse  dynamic susceptibility $\chi^{''}(q)$ through $1/T_{1}T\propto \sum_{q} A(q) \frac{\chi^{''}(q)}{2\pi\nu}$, where $A(q)$ is the hyperfine coupling constant and $\nu$ is the NMR or NQR frequency \cite{Moriya}. Thus, the observed $1/T_{1}T$ behavior indicates that spin fluctuations develop below $T\sim$ 100 K.

\begin{figure}[tbp]
\includegraphics[width=9cm]{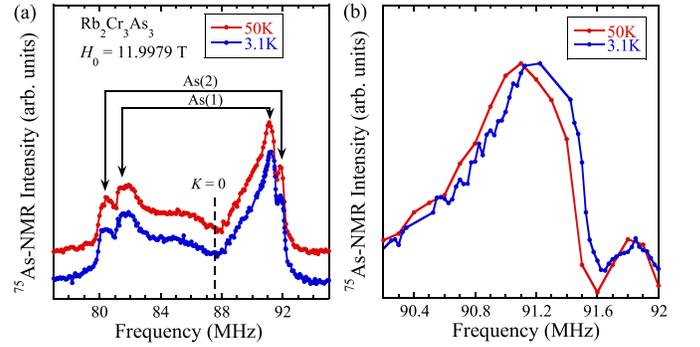}
\centering
\caption{(Color online). (a) $^{75}$As NMR spectra at 50 K and 3.1 K. Each site of As has two peaks with $\theta$ = 42$^{\circ}$ (around 81 MHz) and  $\theta$ = 90$^{\circ}$ (around 91 MHz). (b) The enlarged part of the peak with $\theta$ = 90$^{\circ}$.
\label{NMR}}
\end{figure}

In order to gain insight into the nature of the spin fluctuations, we have conducted  $^{75}$As NMR measurements. 
Figure~\ref{NMR} shows two representative frequency-swept NMR spectra for the central transition ($I_z$ = -1/2 $\leftrightarrow$  +1/2), 
which is  perturbed  by the nuclear  quadrupole interaction.  The   central transition frequency $\nu_{res}$ to the second order  is
given by \cite{Abragam}
\begin{equation}
\nu_{res} = (1+K)\gamma H_{0} + \frac{{3\nu _Q^2}}{{16(1 + K){{(\gamma H_{0})}}}} sin^2 \theta (1 - 9 cos^2 \theta)
\label{eq1}
\end{equation}
where $K$ is the Knight shift due to the hyperfine coupling, and $\theta$ is the angle between 
$H_{0}$ and the principle axis of the electric field gradient which is presumably the crystalline $c$-axis in the present case.
The observed spectrum is in agreement with the theoretically-expected characteristic powder pattern with two peaks at $\theta$ = 42$^{\circ}$ 
and $\theta$ = 90$^{\circ}$. 
For a randomly-oriented powder sample, the peak at $\theta$ = 42$^\circ$ would have a larger intensity than that at $\theta$ = 90$^\circ$. Our results suggest that the powder is partially oriented. 
With the knowledge of 
the  $\nu_{Q}$ value 
obtained from NQR, we can assign the four peaks  into two sets, as marked by the arrows in Fig.~\ref{NMR}. The inner set comes from the As(1) site with smaller $\nu_{Q}$, and the outer set from the As(2) site with larger $\nu_{Q}$. 


\begin{figure}[tbp]
\includegraphics[width=8cm]{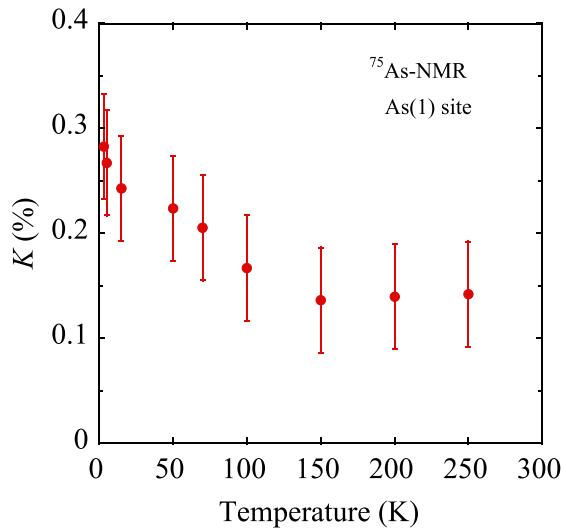}
\caption{(Color online) The $T$-dependence of the $^{75}$As  Knight shift $K$ at As(1) site in the normal state. 
 The error bar was estimated from the
frequency interval in measuring the NMR spectra.
\label{K}}
\end{figure}


We focus on the peak with  $\theta$ = 90$^{\circ}$, and calculated the $K$ according to eq. (1). 
The $K$ result for As(1) site is shown in Fig.~\ref{K}, which is
almost  $T$-independent above $T$ = 150 K. Below $T$ = 100 K,  however, $K$  increases with decreasing $T$. Such increase of $K$  can be appreciated in the change of the NMR spectra as shown in Fig.~\ref{NMR}(b). The shift of the spectrum between $T$ = 3.1 K (where superconductivity was suppressed by the magnetic field of 12 T) and 50 K is about 80 kHz. However, the change of $\nu_Q$ in this $T$-range is small as shown in Fig.~\ref{NQR}, whose contribution (the third term of eq. (1)) is less than 25 kHz.  Therefore, the shift of the spectrum shown in Fig.~\ref{NMR}(b) is predominantly due to the increase of $K$.
%
It is noted that the $T$-dependence of $K$ is similar to that of  1/$T_{1}T$. 
This result is in sharp contrast to  the cobalt-oxide superconductor Na$_x$CoO$_2$$\cdot$1.3H$_2$O with antiferromagnetic fluctuations, where  1/$T_{1}T$ also increases upon cooling~\cite{Fujimoto,zheng} but the Knight shift is $T$-independent  above $T_{\text{c}}\sim$4.3 K up to 100 K~\cite{zheng_PRB}.

Our results for Rb$_{2}$Cr$_{3}$As$_{3}$ are consistent with 3D  ferromagnetic spin fluctuation associated with the 3D Fermi surface which has the predominant contribution to DOS at the Fermi level.
According to Moriya's theory for a  ferromagnetically correlated 3D metal \cite{Moriya}, 1/$T_{1}T$ and $K$ should be proportional to $\chi (\emph{\textbf{q}} = 0)$ which follow a Curie-Weiss $T$-dependence. Indeed, our data can be fitted by $1/T_{1}T = a + b/(T+\theta_{\rm C})$ where the $a$-term  is due to the DOS at the Fermi level ($N_{0}$), 
and the second term is the contribution from the 3D ferromagnetic spin fluctuation. Note that the constant term was not considered in the analysis   for K$_{2}$Cr$_{3}$As$_{3}$, where the authors attributed the relaxation in the whole $T$-range to  1D spin excitations as in  spin chains \cite{Imai}. However,  this term is important for a metal like the present case.
In fact, band calculation for K$_{2}$Cr$_{3}$As$_{3}$ shows that the 3D $\gamma$-band has a quite large contribution to the DOS; 
 the $N_{0}$ from $\gamma$-band, quasi-1D $\alpha$ and $\beta$ bands is respectively 11.55, 1.29 and 2.68 states/(eV$\cdot$ unit cell)  \cite{private}. 
As seen in Fig. 2, the constant term amounts to 1/3 of the observed  $1/T_{1}T$ just above $T_{\rm c}$, and a power-law cannot fit our data. 

\begin{figure}[tbp]
\includegraphics[width=8cm]{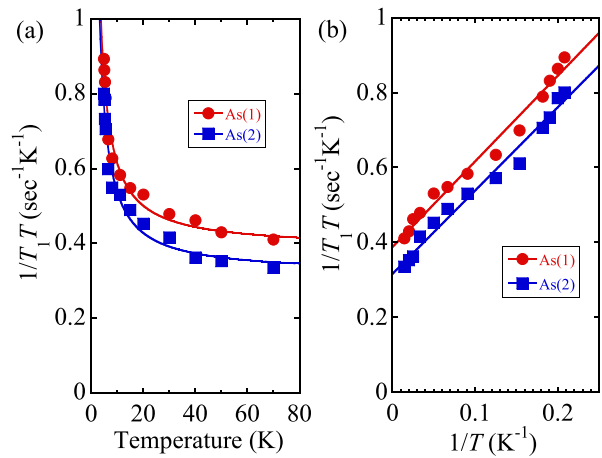}
\caption{(Color online). (a) An enlarged part of $1/T_1T$ for $T_c\leq T <$  80 K. The solid curves are fittings by the Curie-Weiss law, $1/T_{1}T = a + b/(T+\theta_{\rm C})$, with $\theta_{\rm C} \sim$ 0 K(see text). (b) A plot of the same data as (a) to emphasize the low-$T$ part.
 \label{T1T_fit}}
\end{figure}

\begin{figure}[tbp]
\includegraphics[width=8cm]{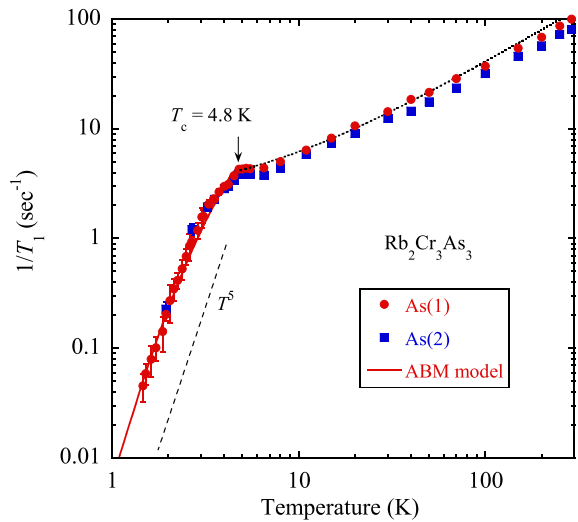}
\caption{(Color online). The $T$-dependence of  1/$T_{1}$ at zero magnetic field. The arrow indicates $T_{\text{c}}$ and the straight line indicates $1/T_1\propto T^5$. The curve below $T_{\rm c}$ is a calculation of the ABM gap with $\Delta_0$=2.1$k_{\rm B}T_{\rm c}$ and $\delta$=$\frac{\Delta_0}{15}$, where $\delta$ is the broadening width of the energy level proposed by Hebel \cite{Hebel} to account for the effects such as impurity scattering \cite{Maclaughlin-Review}. 
The dotted curve above $T_{\rm c}$ is the same result as the curve in Fig. \ref{T1T_fit}. A slight deviation from the data points  is seen at high-$T$, which may indicate that  the 1D Fermi sheets can play some role at high-$T$ \cite{Supple}.
 \label{T1}}
\end{figure}

As shown in  Fig. ~\ref{T1T_fit} (a),
the fitting is fairly good, 
with  $\theta_{\rm C} \sim$  0 K.
A plot emphasizing the low-$T$  part  is shown in Fig. 5(b).
The feature of small $\theta_{\rm C} \sim$  0 K   can be seen more intuitively and directly in the plot of $1/T_1$ vs $T$ as shown in Fig.~\ref{T1}. One notes that $1/T_{1}$ becomes almost $T$-independent before superconductivity sets in, which is an important feature  worth of emphasizing. It follows  from the Moriya theory for $1/T_1T$ that  $\theta_{\rm C}$= 0 K.
This is a remarkable result, which suggests that Rb$_{2}$Cr$_{3}$As$_{3}$ is very close to a ferromagnetic QCP where $\theta_{\rm C}$ = 0 K.
In passing, we remark that no $T$-range over which $1/T_{1}$ is constant was found in K$_{2}$Cr$_{3}$As$_{3}$ \cite{Imai}. One possible reason for this is that the K-compound is a little far away from a QCP, as theory suggested \cite{Hu_jiangping}. 

Next we turn to the properties of the superconducting state. As seen in Fig.~\ref{T1}, 
$1/T_{1}$ drops rapidly below  $T_{\text{c}}$ = 4.8 K,  with  no Hebel-Slichter coherence peak, suggesting unconventional nature of the superconductivity. 
The $1/T_{1}$ in the superconducting state is expressed as \cite{Maclaughlin-Review,Ueda}
\begin{eqnarray}
\frac{T_{1N}}{T_{1S}}&=&\frac{2}{k_BT}
\int\left(1+\frac{\Delta^2}{EE'}\right)N_S(E)N_S(E')\nonumber\\
&\times&f(E)\left[1-f(E')\right]\delta(E-E')dEdE',
\end{eqnarray}
where $1/T_{1N}$ ($1/T_{1S}$) is the relaxation rate in the normal (superconducting) state,   $N_S(E)=N_0 E/\sqrt{E^2-\Delta^2}$ is the DOS in the superconducting state,
 $f(E)$ is the Fermi distribution function and $C =1+ \frac{\Delta^2}{EE'}$ is the coherence factor.
For an $s$-wave gap, the coherence factor and the divergence of the DOS at $E=\Delta$ will lead to a Hebel-Slichter peak  just below $T_{\rm c}$. If there are nodes in the gap function, however, the term $\frac{\Delta^2}{EE'}$ 
disappears due to sign change, then the  coherence peak will not be observed, as  in high-$T_{\rm c}$ cuprates \cite{Asayama}.

Most remarkably, our  $1/T_{1}$ follows a $T^5$ variation at low temperatures.
This result is different from 
 the previous measurement   for K$_{2}$Cr$_{3}$As$_{3}$ where $1/T_1$ was reported to follow a $T^4$-variation \cite{Imai}. 
 Most plausibly, the difference arises from the different procedure in obtaining $1/T_1$. For K$_{2}$Cr$_{3}$As$_{3}$,    the $1/T_1$ value below $T_{\rm c}$ was obtained by a stretched exponential fitting
 , which is different from our procedure using a single exponential fit.
At the moment, we cannot rule out the possibility that the gap function is  different between the two compounds although we are not aware of any gap function that can lead to a $T^n$-variation of $1/T_1$ with an even number of $n$ (see below for detail).

The important feature of $1/T_1\propto T^5$ is consistent with the existence of point nodes in the gap function. For example, in the Anderson-Brinkman-Morel (ABM) model  \cite{ABM,ABM2},  
 the gap $\Delta = \Delta_0 sin \theta$e$^{i\phi}$ has two 
point nodes at $\theta$ = 0 and $\pi$. In such case,  $N_S(E)\propto E^2$ at low-$E$, which results in a $T^5$-variation of $1/T_1$ following eq. (2). Experimentally, a $T^5$-variation was  reported so far only in the heavy-fermion superconductor PrOs$_4$Sb$_{12}$ under pressure \cite{Katayama}. As a comparison,  for a gap with line nodes, $N_S(E)\propto E$ at low-$E$ and then a $T^3$-variation is seen below $T_{\rm c}$ as in some Ce-based heavy fermion compounds \cite{Zheng-115}.
The curve below $T_{\rm c}$ in Fig. \ref{T1} is a calculation using the ABM model with
$\Delta_0$=2.1$k_{\rm B}T_{\rm c}$.  The agreement between the theoretical curve and the data points is good at low temperatures but the data points deviate from the curve at $T\sim$ 3 K.

The  feature that $1/T_1$  changes slop at $T\sim$ 3 K resembles that in the Fe-pnictides superconductors \cite{Matano,Li}, where multiple gaps opened on different Fermi surfaces.  Our result may be an indication of multiple gaps since the present compound is also a multiple-bands system \cite{note}. For a two-gap system, the physical quantities just below $T_{\rm c}$ are dominantly governed by the larger gap.
Only at low temperatures where the thermal energy
becomes comparable to or smaller than the smaller gap, the system realizes the existence of the smaller gap, resulting in another drop of the quantity like  $1/T_1$ \cite{Matano}.
Recently, many superconductors including heavy fermion compounds have been reported to exhibit multiple-gaps features. For example,   the specific heat measurement on UBe$_{13}$ with multiple Fermi surfaces has revealed three gaps with different size \cite{Sakakibara}. However, the multiple-gap feature only appeared at  $T=0.1\sim0.2T_{\rm c}$  so that it was not evident in  $1/T_{1}$ which is usually affected by impurity at such low $T$ \cite{MacLaughlin}. Unfortunately, the simple ABM model with two gaps is unable to satisfactorily fit our data \cite{Supple},  and we  call for more theoretical investigations for the gap function pertinent to the present compound.


%
%


In summary, we found strong spin fluctuations in Rb$_{2}$Cr$_{3}$As$_{3}$ by the measurements of $^{75}$As NQR spin-lattice relaxation rate 1/$T_{1}$ and the Knight shift  $K$. Both  $K$ and  1/$T_{1}T$    increase upon cooling below 100 K, which are consistent with   ferromagnetic spin fluctuation. In the superconducting state, the lack of Hebel-Slichter coherence peak below $T_{\text{c}}$ and the $T^5$ variation of the spin-lattice relaxation rate 1/$T_{1}$ at low temperatures   suggest  unconventional superconductivity with point nodes in the gap function. Our results provide new insight into unconventional superconductivity in strongly-correlated electron systems.

\begin{acknowledgments}
We thank C. Cao, F. Yang, J.P. Hu and S. Kawasaki for useful discussions, and Z. Li for assistance.
This work
was partially supported by CAS Strategic Priority
Research Program No. XDB07020200,  National Basic Research Program of China (Nos. 2011CBA00109 and 2012CB821402) and NSFC Grant No 11204362.

\end{acknowledgments}


\end{document}